\documentclass[12pt]{article}
\usepackage{graphicx}
\usepackage{subfigure}
\input epsf.tex

\newcommand{\beq}{\begin{equation}}
\newcommand{\eeq}{\end{equation}}
\newcommand{\be}{\begin{equation}}
\newcommand{\ee}{\end{equation}}
\newcommand{\bea}{\begin{eqnarray}}
\newcommand{\eea}{\end{eqnarray}}
\newcommand{\req}[1]{~(\ref{#1})}

\newcommand{\gs}{\mbox{$g_s$}}            
\newcommand{\ap}{\mbox{$\alpha^\prime$}}  
\newcommand{\ls}{\mbox{$l_s$}}            

\def\href#1#2{#2}

\def\p{\partial}

\begin{document}

\baselineskip=15.5pt
\pagestyle{plain}
\setcounter{page}{1}

\begin{titlepage}
\begin{flushleft}
       \hfill                      {\tt hep-th/1003.3698}\\
       \hfill                       FIT HE - 10-01 \\
       \hfill                       KYUSHU-HET 124 \\
       \hfill                       Kagoshima HE - 10-1 \\
\end{flushleft}

\begin{center}
  {\huge Holographic Penta and Hepta Quark State \\ 
\vspace*{2mm}
in Confining Gauge Theories}
\end{center}

\begin{center}

\vspace*{2mm}
{\large Kazuo Ghoroku${}^{\dagger}$\footnote[1]{\tt gouroku@dontaku.fit.ac.jp},${}^{\S}$Akihiro Nakamura\footnote[3]{\tt nakamura@sci.kagoshima-u.ac.jp}
Tomoki Taminato${}^{\ddagger}$\footnote[2]{\tt taminato@higgs.phys.kyushu-u.ac.jp},
\\
and ${}^{\P}$Fumihiko Toyoda\footnote[4]{\tt ftoyoda@fuk.kindai.ac.jp}
%
}\\

\vspace*{4mm}
{${}^{\dagger}$Fukuoka Institute of Technology, Wajiro, 
Higashi-ku} \\
{
Fukuoka 811-0295, Japan\\}
{
${}^{\ddagger}$Department of Physics, Kyushu University, Hakozaki,
Higashi-ku}\\
{
Fukuoka 812-8581, Japan\\}
{
${}^{\S}$Department of Physics, Kagoshima University, Korimoto 1-21-35, \\Kagoshima 890-0065, Japan\\}
{
${}^{\P}$School of Humanity-Oriented Science and
Engineering, Kinki University,\\ Iizuka 820-8555, Japan}

\vspace*{5mm}
\end{center}

\begin{center}
{\large Abstract}
\end{center}
We study a new embedding solutions of D5 brane in an 
asymptotic AdS${}_5\times S^5$ space-time, which is
dual to a confining $SU(N_c)$ gauge theory. The D5 brane
is wrapped on $S^5$ 
as in the case of the vertex of holographic baryon. However, the solution 
given here is different from the usual baryon vertex in the point that it couples 
to $k$-anti-quarks 
and $N_c+k$ quarks on the opposite two points of $S^5$, the north and south poles,
respectively. 
The total quark number of this state is preserved as $N_c$ when minus one 
is assigned to anti-quark, then it forms a color singlet like the baryon. 
However, this includes anti-quarks and quarks, whose
number is larger than that of the baryon. When we set as $N_c=3$,
we find the so called
penta and hepta-quark states. 
We study the dynamical properties of these states by solving the vertex and
string configurations for such states. 
The mass spectra of these states and the tension of the stretched vertex
are estimated, and they are compared with that of the baryon.

\noindent

\vfill
\begin{flushleft}

\end{flushleft}
\end{titlepage}
\newpage

\section{Introduction}

In the context of string/gauge theory correspondence
\cite{jthroat,gkpads,wittenholo}, 
the baryon has been studied as a system of fundamental strings (F-strings) and 
D5 branes wrapped on $S^5$ in AdS${}_5\times S^5$ space-time 
\cite{wittenbaryon, groguri,imamura,cgs,GRST,cgst,ima-04,baryonsugra,imafirst}.
They are dual to the quarks and the baryon vertex respectively.
The F-strings dissolve as the $U(1)$ flux in the D5 brane, and 
they flow out as separated $N_c$ free strings from the cusp(s) on  
the D5 brane. 

This idea has been recently studied furthermore
\cite{GI} along the approach given in \cite{imamura}-\cite{cgst},
and also extended to the finite temperature case \cite{GINT,BCH,ALR}. 
We could find complicated structures of the D5 brane embedded 
in the background dual to a confining gauge theory.
Especially, a new 
configuration has been found and proposed as the baryonium, which is constructed
of $k$-quarks, $k$-anti-quarks and the vertex (D5 brane)  \cite{GINT2,GINT3}.
In this case, the vertex is described by the polar angle ($\theta$) on $S^5$
in the range $0<\theta_0\leq \theta\leq\pi$, and the D5 brane solution is
two valued for $\theta$. 

In the case of the baryon,
the D5 brane 
covers the range $0\leq \theta\leq\pi$. 
The embedded solution has cusps at 
the pole points, $\theta=0$ and $\pi$, where the quarks are attached.
Their total 
number is $N_c$ to form a color singlet. The partition 
of the quark numbers on the two poles is determined
by a constant parameter $\nu$ ($0\leq \nu\leq 1$), which is obtained as an integration
constant of the equation of motion for the D5 brane vertex. Namely they are
separated as $N_c\nu$ and $N_c(1-\nu)$. These numbers should be quantized as 
integers physically.

Here, we propose a new configuration which is given by extending the 
parameter $\nu$ to the negative region $\nu < 0$. 
This extension is possible since the parameter $\nu$ is
an arbitrary integration constant in solving the equations of motion.
Then we find a new hadron configuration, which has 
$N_c|\nu|$ anti-quarks 
at one cusp and $N_c(1+|\nu|)$ quarks on the other one.
While the total number of the quarks is preserved as $N_c$ when the anti-quark
number is assigned as minus, 
but this configuration is clearly different from the usual baryon.

It is possible to have the same quark and anti-quark 
numbers in different configurations which are
obtained by attaching some pairs of quark and anti-quark on the
same cusp point 
of $S^5$. However such pairs would be removed by considering 
the pair annihilation of the 
quark and anti-quark on the D5 brane. 
When we remove the anti-quark
by this rule, we find configurations, in which the quarks and 
the anti-quarks are separated to the opposite side of $S^5$ each other 
as stated above. In this case, the anti-quarks cannot be removed 
any more by the rule of pair annihilation since they are separated.

\begin{figure}[htbp]
\vspace{.3cm}
\begin{center}
 \includegraphics[width=14cm,height=7cm]{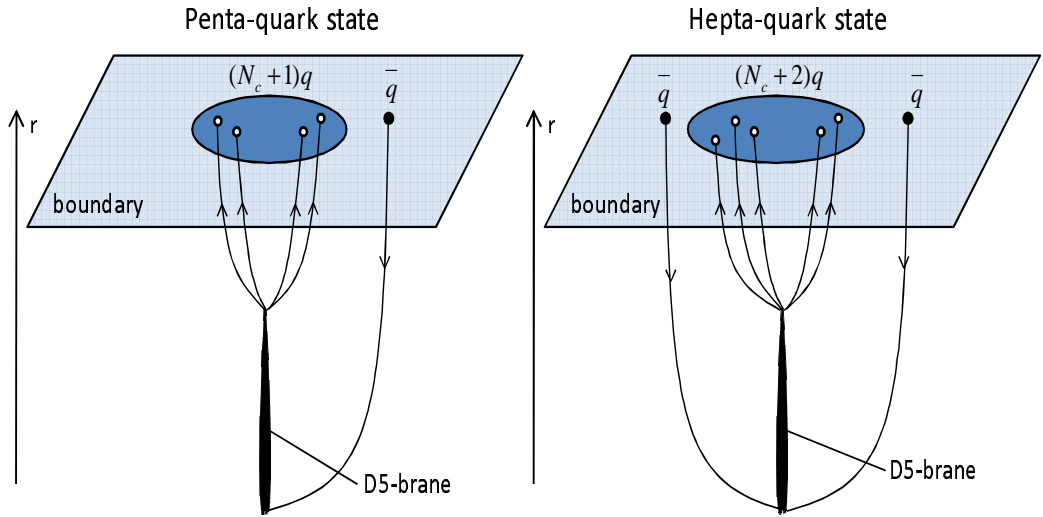}
\caption{\small The configuration of penta and hepta-quark states
with a D5 brane vertex for $N_c=3$ theory.
\label{penta-pic-1}}
\end{center}
\end{figure}

Here we concentrate on these new solutions of negative $\nu$.
As a simple example, consider the case of $N_c=3$ or equivalently
$SU(3)_c$, then we find a configuration
of one anti-quark at one cusp and four quarks on the other one. 
The corresponding configuration is shown in the Fig.\ref{penta-pic-1}.
This might
be considered as the candidate of the
penta-quark \cite{Nakano}. In this case, the anti-quark is separated
from the quarks on the opposite poles of 
the D5 brane, then they are stable against the pair annihilation
as stated above. 
In the Fig.\ref{penta-pic-1}, such
possible configurations, the penta and hepta quark
states, are shown for $N_c=3$ case.
However, we find that there is no stable solution
for the penta quark, and the stable states
are found only for the case 
$N_c\nu\leq -2$. The stability is assured by the balance conditions of the force
at each cusp. For penta quark, the condition at the lower cusp-point is not
satisfied as shown below.

The energy and the configuration
of those exotic states depend on the configuration of the state.
It is determined by the boundary conditions of the equations
of motion, which are called as no force conditions. 
So, varying the boundary conditions, 
the relation
between the vertex energy and the distance of the two cusps is examined.
Then we could estimate the tension of the vertex, and we could find a 
minimum energy configuration of the hepta quark. Its ratio to the baryon
mass is given by $7/3$, which is equivalent to the one of the quark number.

In Section \ref{eqnsec} we give our model and D5 brane action
with non-trivial $U(1)$ gauge field. 
And the equations of motion for D5 brane are given, and 
various kinds of solutions are explained.
In section
\ref{bound}, the quarks and the vertex with negative $\nu$ are studied
by solving the equations of motion with no force conditions.
In the section 4, the stable hepta quark states are studied 
numerically for $N_c=3$.
In the section 5, the new baryonic states are 
examined in other holographic models for the confining gauge theories,
and we find a common property to forbid the penta quark.
And in the final section, we summarize our results and discuss related
problem.

\section{Model}\label{eqnsec}

\subsection{Bulk background}

We start from
10d IIB model retaining the dilaton
$\Phi$, axion $\chi$ and self-dual five form field strength $F_{(5)}$
with the following action, 
\beq\label{10d-action}
 S={1\over 2\kappa^2}\int d^{10}x\sqrt{-g}\left( {\cal R}-
{1\over 2}(\partial \Phi)^2+{1\over 2}e^{2\Phi}(\partial \chi)^2
+{1\over 4\cdot 5!}F_{(5)}^2
\right), \label{5d-action}
\eeq
where the axion is Wick rotated to obtain the solution given in \cite{Liu:1999fc}.
We notice that
the axion $\chi$ corresponds to the source of D(-1) brane and it is Wick rotated
in the supergravity action. This is necessary to preserve the supersymmetry.
Under the Freund-Rubin
ansatz for $F_{(5)}$, 
$F_{\mu_1\cdots\mu_5}=-\sqrt{\Lambda}/2~\epsilon_{\mu_1\cdots\mu_5}$ 
 \cite{Liu:1999fc,KS2,GY}, and for the 10d metric as $M_5\times S^5$ or
$ds^2=g_{mn}dx^mdx^n+g_{ij}dx^idx^j$, the equations of motion
are solved. 
\footnote{The five dimensional $M_5$ part of the
solution is obtained by solving the following reduced 5d action,
$$ 
 S={1\over 2\kappa_{(5)}^2}\int d^5x\sqrt{-g}\left( {\cal R}_{(5)}+3\Lambda-
{1\over 2}(\partial \Phi )^2+{1\over 2}e^{2\Phi}(\partial \chi )^2
\right), $$ 
which is written 
in the string frame and taking $\alpha'=g_s=1$. }
Here $m,~n=0\sim 4$, $i,~j=5\sim 9$ and $\Lambda$ denotes a constant.  

\vspace{.3cm}
The solution is obtained under the ansatz,
\beq
\chi=-e^{-\Phi}+\chi_0 \ ,
\label{super}
\eeq
where $\chi_0$ is an arbitrary constant. This ansatz 
is necessary to obtain supersymmetric solutions, and 
\bea\label{background}
ds^2_{10}&=&G_{MN}dX^{M}dX^{N}\nonumber \\
&=&e^{\Phi/2}
\left\{
{r^2 \over R^2}A^2(r)\left(-dt^2+\sum_{i=1}^{3}(dx^i)^2\right)+
\frac{R^2}{r^2} dr^2+R^2 d\Omega_5^2 \right\}  . 
\label{finite-c-sol}
\eea 
Then, the supersymmetric solution is obtained as
\beq
e^\Phi= 1+\frac{q}{r^4}\ , \quad A=1\, ,
\label{dilaton}
\eeq
where $M,~N=0\sim 9$,
$R=\sqrt{\Lambda}/2=(4 \pi N)^{1/4}$ and $N_c$ denotes the number of D3 branes.
The parameter $q$ represents the vacuum expectation value (VEV) 
of gauge fields condensate~\cite{GY}. 
In this configuration, the four dimensional boundary represents the 
$\cal{N}$=2 SYM theory. In this model, we find quark confinement in the
sense that we find a linear rising potential between quark and anti-quark
with the tension $\sqrt{q}/R^2$ \cite{KS2,GY}.


\vspace{.3cm}
Then, we solve the embedding equations of D5 brane and fundamental string 
in the following 10D background, 
\beq
ds^2_{10}= e^{\Phi/2}
\left(
\frac{r^2}{R^2}\eta_{\mu\nu}dx^\mu dx^\nu +
\frac{R^2}{r^2} dr^2+R^2 d\Omega_5^2 \right) \ ,
\label{SUSY-metric}
\eeq
where $e^\Phi$ is given above. We notice the equivalence of
this metric in the Einstein frame with the one of the AdS$_5\times S^5$.

\subsection{D5 brane}
\label{eqnsec2}
The baryon is constructed from the vertex and $N$
fundamental strings~ \footnote{
Hereafter we denote $N_c$ as $N$ for simplicity.}, and
the vertex is given by the D5 brane wrapped on the
${S}^{5}$ of the above metric. 
The $N$ fundamental strings terminate on this vertex and they are dissolved
in it \cite{wittenbaryon,groguri} as $U(1)$ flux. 
The D5 brane action is thus written by the Dirac-Born-Infeld (DBI) plus
WZW term \cite{cgs}
\begin{eqnarray}\label{d5action}
S_{D5}&=&-T_{5}\int d^6\xi
 e^{-\Phi}\sqrt{-\det\left(g_{ab}+\tilde{F}_{ab}\right)}+T_{5}\int
d^6\xi \tilde{A}_{(1)}\wedge {\cal G}_{(5)} ~,\\
g_{ab}&\equiv&\p_a X^{M}\p_b X^{N}G_{MN}~, \qquad
{\cal G}_{a_1\ldots
a_5}\,\equiv\,\p_{a_1}X^{M_1}\ldots\p_{a_5}X^{M_5}G_{M_1\ldots M_5}~.\nonumber
\end{eqnarray}
where $\tilde{F}_{ab}=2\pi\ap F_{ab}$ and 
$T_5=1/(\gs(2\pi)^{5}\ls^{6})$ is the brane tension. And ${\cal G}_{(5)}$ represents
the induced five form field strength, which is obtained from the bulk five form,
\begin{equation}\label{fiveform}
G_{(5)}\equiv dC_{(4)}=4R^4\left(\mbox{vol}(S^5) d\theta_1\wedge\ldots\wedge d\theta_5
-{r^3\over R^8}  dt\wedge \ldots\wedge dx_3\wedge dr\right),
\end{equation}
where $\mbox{vol}(S^5)\equiv\sin^4\theta_1
\mbox{vol}(S^4)\equiv\sin^4\theta_1\sin^3\theta_2\sin^2\theta_3\sin\theta_4$.

\vspace{.3cm}
The D5 brane is embedded in the world volume
$\xi^{a}=(t,\theta,\theta_2,\ldots,\theta_5)$, where $(\theta_2,\ldots,\theta_5)$
are the $S^4$ part with the volume, $\Omega_{4}=8\pi^{2}/3$, where we set
as $\theta_1=\theta$. 
Restrict our attention to $SO(5)$ symmetric configurations of the 
form $r(\theta)$, $x(\theta)$, and $A_t(\theta)$ (with all other fields 
set to zero). 
Then the above action is written as
\be \label{d3action}
S= T_5 \Omega_{4}R^4\int dt\,d\theta \{ -\sin^4\theta 
  \sqrt{e^{\Phi}\left(r^2+r^{\prime 2}+(r/R)^{4}x^{\prime 2}\right)
   -\tilde{F}_{t \theta}^2}  -\tilde{F}_{t \theta} D \},
\ee
where the WZW term is rewritten by partial integration with respect to
$\theta$. 
The factor $D(\theta)$ is therefore defined by
\beq\label{D3}
\partial_\theta D = -4 \sin^4\theta~,
\eeq
then it is solved as
\be \label{d}
  D(\nu,\theta) \equiv \left[{3\over 2}(\nu\pi-\theta)
  +{3\over 2}\sin\theta\cos\theta+\sin^{3}\theta\cos\theta\right]\, ,
\ee
where the integration constant $\nu$ plays an important role
to determine the configuration of the D5 brane, and its meaning 
is explained below.

Before considering $\nu$, we rewrite the above action by eliminating
$\tilde{F}_{t \theta}$ in terms of the Legendre transformation. From
(\ref{d3action}), the equation of motion for $\tilde{A}_{t}$ is obtained
as
\beq
\partial_{\theta}\left({\sin^4\theta~ \tilde{F}_{t \theta}\over
  \sqrt{e^{\Phi}\left(r^2+r^{\prime 2}+(r/R)^{4}x^{\prime 2}\right)
   -\tilde{F}_{t \theta}^2}}-D\right)=0\, ,
\eeq
where 
$r'=\partial_{\theta}r(\theta)$ and $x'=\partial_{\theta}x(\theta)$.
This equation is solved as
\beq\label{D2}
  D={\sin^4\theta~ \tilde{F}_{t \theta}\over
  \sqrt{e^{\Phi}\left(r^2+r^{\prime 2}+(r/R)^{4}x^{\prime 2}\right)
   -\tilde{F}_{t \theta}^2}}~+c_1.
\eeq
Here a new $\theta$-independent constant ($c_1$) seems to be appeared. However,
this is absorbed by the constant $\nu$ given in Eq.~(\ref{d}), then we set
this constant to be zero without any ambiguity. From the viewpoint of
$U(1)$ gauge theory, we can
call this $D$ as displacement, and is related to $\tilde{F}_{t \theta}$ 
by the above  equation with $c_1=0$.

Then, the action is rewritten by the Legendre transformation with respect to
the field $\tilde{A}_{t}$, and we 
obtain an energy
functional written by the embedding coordinate only
\footnote{
$U$ is obtained by a Legendre transformation of $L$, which is defined
as $S=\int dt L$, as $U={\partial L\over \partial\tilde{F}_{t \theta}}
\tilde{F}_{t \theta}-L$. Then equations of motion of (\ref{d3action}) 
provides the same solutions of the one of $U$. 
}
\be \label{u}
U = {N\over 3\pi^2\alpha'}\int d\theta~e^{\Phi/2}
\sqrt{r^2+r^{\prime 2} +(r/R)^{4}x^{\prime 2}}\,
\sqrt{V_{\nu}(\theta)}~.
\ee
\be\label{PotentialV}
V_{\nu}(\theta)=D(\nu,\theta)^2+\sin^8\theta
\ee
where we used $T_5 \Omega_{4}R^4=N/(3\pi^2\alpha')$.
\footnote{Here $\alpha'$ is retained for reminding the physical dimension.}
Then, $r(\theta)$ and $x(\theta)$
are obtained by the variation of this energy function $U$ for the embedding solutions
of the D5 brane. Before solving the equations of motion, we give comments on the
equations obtained from (\ref{d3action}).

\vspace{.23cm}
\noindent{\bf Equations of motion for $r(\theta)$ and $x(\theta)$ from Eq.(\ref{d3action})}
\vspace{.3cm}

The vertex part of the penta-quark state is given by solving the
D5 brane action (\ref{d3action}) or its energy function (\ref{u}).
The equations of the latter case are useful. Then they have been used
in \cite{cgs,GI,GINT,GINT2} and are shown in the Appendix. We use them
here also.

Of course, we can obtain the same results when we start from (\ref{d3action}).
Here we comment on the relation between the equations obtained from (\ref{d3action})
and the one from (\ref{u}).
The equation for $A_t$ is already given above, then we write other equations
for $r(\theta)$ and $x(\theta)$, they are given as follows
\bea\label{classical-1}
    \sin^4\theta\partial_r\left(e^{\Phi/2} \sqrt{\tilde{K}_{(0)}}\right)
     -\partial_{\theta}\left(\sin^4\theta {e^{\Phi/2}\over \sqrt{\tilde{K}_{(0)}}} r'\right)
         &=&0\, , \nonumber \\
    \partial_{\theta}\left(\sin^4\theta {r^4 e^{\Phi/2}\over R^4 \sqrt{\tilde{K}_{(0)}}} x'\right)
         &=&0\, ,  
\eea 
where 
\beq
  \tilde{K}_{(0)}=\left(r^2+(r')^2+{r^4\over R^4} (x')^2-e^{-\Phi}\tilde{F}_{t \theta}^2\right)\, .
\eeq 
The latter equation of (\ref{classical-1}) is solved as
\beq\label{classical-2}
    \sin^4\theta {r^4 e^{\Phi/2}\over R^4 \sqrt{\tilde{K}_{(0)}}} x'
         =h\, ,
\eeq
where $h$ is a constant representing a conserved quantity due to the translation invariance
in the direction of $x$ in $U$.
This is identified with the one used in our previous paper 
by the same notation \cite{GINT2}. 
Actually, the parameter $h$ given in the above
equations (\ref{classical-1}) 
is equivalent with the one of equation (\ref{Heom})
in the Appendix.
Using this $h$ and $A_t$ solved above, 
we obtain the equation for $r(\theta)$. Then we obtain the solutions
by solving this equation, however, it will be solved
numerically since it has very complicated form.  

\vspace{.3cm}
It is easy to prove the equivalence of the solutions of the above
equations (\ref{classical-1}) and the one given in \cite{GINT2} in more efficient way
as given in the Appendix.

\vspace{.8cm}
\noindent {\bf Meaning of $\nu$ and various solutions }

The above equations are common to all the vertex configurations of 
baryon \cite{GI,cgs}, baryonium \cite{GINT2,GINT3} and new baryonic 
stats. They are discriminated by the value of $\nu$ and the range of 
variable $\theta$.

\vspace{.3cm}
\noindent {\bf (i) Baryon: $0\leq \theta\leq \pi$, $0\leq\nu\leq 1$ }

First, we consider the value of $\nu$ defined in the range of 
$0\leq\nu\leq 1$.  This case corresponds to the baryon.  Using 
(\ref{u}), we review the meaning of the integration constant $\nu$ in 
(\ref{d}).  The solution has two cusps at $r(\theta=\pi)$ and $r(0)$, 
namely at poles on $S^5$, $\theta=0$ and $\theta=\pi$.  At these points, 
$r'=\partial_{\theta}r\to \infty$ and $x'\simeq 0$ for $q\to 0$ as shown 
in \cite{GI,cgs}.  Then the configuration near these positions represents 
the bundle of the fundamental strings, and their numbers at the cusps are 
estimated as follows. At $\theta=\pi$ and for $q\to 0$ ($\Phi \to 0$), we 
obtain the following approximate formula
\beq
U \simeq {N\over 3\pi^2\alpha'}\int dr~{3\over 2}(1-\nu)\pi~
={N\over 2\pi\alpha'}(1-\nu)\int dr~.
\label{upi}
\eeq
And similarly, we obtain the following at $\theta=0$,  
\beq
U \simeq {N\over 2\pi\alpha'}\nu\int dr~.
\label{0}
\eeq
Since ${1\over 2\pi\alpha'}\int dr$ represents the bundle of a fundamental 
string, the total number of fundamental strings is given by $N$, which
are separated to $N(1-\nu)$ and $N\nu$ to each cusp point. The meaning of
$\nu$ is then the ratio of this partition, so it must be defined as 
$0\leq\nu(\equiv k/N)\leq 1$, where $k(\leq N)$ is an integer.
Then the total number of the flux is counted as $N$ when we sum up the 
one of the two cusps at $\theta=0$ and $\theta=\pi$.

Next, we give a comment on the orientation of the flux of $U(1)$. It is 
characterized by the sign of the displacement, $D$. 
The function $U$ given by (\ref{u}) is common to positive and negative
$D$. However, at the cusps, $D$ is given as
\beq
  D(\nu,0)={3\over 2}\nu\pi\, , \quad D(\nu,\pi)={3\over 2}(\nu-1)\pi\,
\eeq
and we can see its sign. For $0<\nu<1$, 
we find $D(\nu,0)>0$ and $D(\nu,\pi)<0$. This means that the $U(1)$ flux 
comes into $\theta=0$ then opposite oriented flux does into $\theta=\pi$.
Then we find totally $N$ flux at the baryon vertex. The anti-baryon is
then obtained by the change of $D\to -D$.


\vspace{.3cm}
\noindent{\bf (ii) Baryonium: $\theta_0\leq \theta\leq \pi$}

In general, two possible flux numbers are assigned as $\pm N(1-\nu)$ and
$\pm N\nu$ at each cusp. For the split baryon vertex, which extends between
the cusps at $\theta=\theta_0$, $\pi$, and the flux with the number
$\pm N\nu$ and $\pm N(1-\nu)$ are assigned at each cusps. Then we find 
$\pm N$ quark number since the baryon must be a color singlet. However, 
we found new solutions, which extend between the cusps at the same 
point $\theta(=\theta_0$, $\pi)$ as shown in \cite{GINT2,GINT3}.
In this case, for the solution with two cusps at $\theta=\pi$, we must 
choose the flux-combination as $\pm N(1-\nu)$ and $\mp N(1-\nu)$. And 
for the one with the cusps at $\theta=\theta_0$, the flux should be 
assigned as $\pm N\nu$ and $\mp N\nu$.  Then the total flux or the 
quark number is zero in both cases, and these solutions have been 
assigned as the baryonium states.  

\vspace{.3cm}
\noindent {\bf (iii) New Baryonic States $0\leq \theta\leq \pi$, $\nu<0$
(or $1<\nu$); }

New states are obtained when $\nu$ is set as negative value, $\nu<0$.
In this case, $N|\nu|$ flux comes in $\theta=0$ and $N(1+|\nu|)$ flux
comes out from $\theta=\pi$. This configuration represents a bound state 
of $N|\nu|$ anti-quarks, $N(1+|\nu|)$ quarks and 
the D5 brane as a color singlet. 
As a special case, we can consider a state with $N|\nu|=1$
and $N(1+|\nu|)=4$ for $N=3$. This state is formed with four quarks 
and an anti-quark. And this state is called as 
penta-quarks. 
By interchanging the positions ($\theta=0$ and $\pi$)
of the attached $N|\nu|$ anti-quarks
and the $N|\nu|$ anti-quarks, we find the configurations in the case of $1<\nu$.
It is very interesting to study whether such a state 
exists or doesn't. 
Here we call all the states of $\nu<0$ and $1<\nu$ as 
new baryonic states as a whole. These states have 
not been studied up to now in the holographic model.
Since the states of $\nu<0$ and $1<\nu$ are symmetric, we study the case
of $\nu<0$ hereafter.


\section{New Baryonic States}\label{bound}

As explained above, the new baryonic state is obtained by solving
the equations of the system of D5 brane, quarks and anti-quarks. The
equations of motion of this system are given by the following energy
function,
\beq\label{U-total}
  U_{\rm total}=U_{D5}+U_{\rm F-q}+U_{\rm F-anti-q}
\eeq
where 
\beq
 U_{D5}={N\over 3\pi^2\alpha'}\int_0^{\pi} d\theta~e^{\Phi/2}
\sqrt{r^2+r^{\prime 2} +(r/R)^{4}x^{\prime 2}}
\sqrt{V_{\nu}(\theta)}~\, ,
\eeq
\beq
 U_{\rm F-anti-q}=\sum_{i=1}^{N|\nu|}{1\over{2\pi\alpha'}}
  \int_{r^{(i)}({\theta=0})}^{r_{\rm max}}dr^{(i)}~e^{\Phi/2} 
  \sqrt{1+(r^{(i)}/R)^{4}(x_r^{(i)})^2}\, ,
\eeq
\beq
 U_{\rm F-q}=\sum_{i=1}^{N(1+|\nu|)}{1\over{2\pi\alpha'}}
  \int^{r_{\rm max}}_{r^{(i)}(\theta=\pi)}dr^{(i)}~e^{\Phi/2} 
  \sqrt{1+(r^{(i)}/R)^{4}(x_r^{(i)})^2}\, ,  
\eeq
where $x_r=\partial_r x$.  And $r_{\rm max}$ denotes the cutoff position 
of the F-strings, namely the one of D7 brane.  Each quark or anti-quark 
string is discriminated by its superscript $(i)$.

\vspace{.3cm}
In solving the equations, we notice the following point for the case of
negative $\nu$. In general, $V_{\nu}(\theta)$ has a minimum
at $\theta_c$ which is given as a solution of 
\beq
\pi\nu=\theta_{c}-\sin\theta_{c}\cos\theta_{c}\, ,
\label{nuc}
\eeq
and we find the minimum value as $V_{\nu}(\theta_c)=\sin^6(\theta_c)$.
However the point $\theta_c$ is now negative since $\nu<0$.
Then this point is out of the $S^5$. This implies that the electric field
in the $S^5$ does not vanish at any point of $\theta$, which is restricted
as $0\leq\theta\leq\pi$. So the shape of the embedded D5 brane is largely 
restricted compared to the baryon.

\vspace{.7cm}
{\bf No-force condition}

\vspace{.3cm}
{In the present case, the D5 brane has cusps at the two poles on the $S^5$.
At these points, the tensions of the D5 brane appear toward smearing the
cusp shape. On the other hand, the force coming from this tension of D5 brane
could be balanced by the tension working in the opposite direction when
fundamental strings are added at these cusp points. The condition to balance
both forces are called as no-force conditions and they are studied in
considering baryons \cite{GI, GINT, BLL}. Also in the present case,
we must add quarks and anti-quarks at the cusps of the brane 
to see the full configuration of the penta quark.

The no-force conditions are obtained by considering the boundary terms
of the equations of motion of the system of D5 brane and F-strings
given above.
At the cusp $\theta=0$ ($\theta=\pi$), we give the balance of the 
tensions between the brane and the $|\nu| N$ ($(1+|\nu|) N$) 
fundamental strings corresponding to the anti-quarks (quarks).
They are given for $\theta=0$ as, 
\beq
N|\nu|{r'\over \sqrt{r^2+{r'}^2+(r/R)^4{x'}^2}}
   =\sum_{i=1}^{N|\nu|}{r_x^{(i)}\over \sqrt{(r_x^{(i)})^2+(r/R)^4}}\,.
\label{UFr}
\eeq
for $r$ direction and 
\beq
N|\nu|{x'(r/R)^2\over \sqrt{r^2+{r'}^2+(r/R)^4{x'}^2}}
   =\sum_{i=1}^{N|\nu|}{{x_r^{(i)}(r/R)^2}\over{\sqrt{1+(r/R)^4(x_r^{(i)})^2}}}\,.
\label{UFx}
\eeq
for $x$ direction. For the boundary $\theta=\pi$, we obtain 
the conditions with the same form by replacing
$|\nu| N$ by $(1+|\nu|) N$ in the above equations.
Here the factor $(r/R)^2$ in (\ref{UFr}), (\ref{UFx}) can be removed 
by dividing the left and the right hand sides, but it is retained for 
the purpose given in the following analysis.

In this expression, the above conditions 
are represented by the following two dimensional vectors in the $x-r$ plane,
\bea
 \vec{\tau}_v&=&\left({x'(r/R)^2\over \sqrt{r^2+{r'}^2+(r/R)^4{x'}^2}},~
{r'\over \sqrt{r^2+{r'}^2+(r/R)^4{x'}^2}}\right)\, , \\ 
  \vec{\tau}_{s}^{(i)}&=&\left({{x_r^{(i)}(r/R)^2}\over{\sqrt{1+(r/R)^4(x_r^{(i)})^2}}},
    ~{r_x^{(i)}\over \sqrt{(r_x^{(i)})^2+(r/R)^4}}\right)\, ,
\label{vector-repre} 
\eea
for the vertex ($\vec{\tau}_v$) and strings ($\vec{\tau}_{s}$), respectively.
Here we notice
\beq
 |\vec{\tau}_v|\leq|\vec{\tau}_s^{(i)}|=1\, ,
\label{vector-ineq}
\eeq
where the equality is satisfied for $r=0$.
This is important due to the reason given below. Then the above no-force
conditions are written as
\beq
 N|\nu|\vec{\tau}_v=\sum_{i=1}^{N|\nu|}\vec{\tau}_s^{(i)}\, . 
\label{vector-eq}
\eeq

\vspace{.3cm}
As a special case, consider the configuration of $|\nu| N=1$. In this case,
one fundamental string, which corresponds to an anti-quark, 
is connected at $\theta=0$ to the brane.  For the case of $N=3$, on 
the side of $\theta=\pi$, four F-strings are attached corresponding 
to the four quarks.  Then this configuration corresponds just to the 
so called penta-quark state.  Considering the above condition at 
$\theta=0$, we obtain $|\vec{\tau}_{v}|=|\vec{\tau}_{s}|$ then
\beq
  r^2=0\, .
\eeq
This implies that the point of $\theta=0$ must be on at $r=0$ for the
fundamental quark string. However this configuration needs infinite
energy due to the theory is in the confinement phase \cite{GY}.
Then this configuration, the penta quark state, cannot be realized.

In general, it is possible to satisfy the equality (\ref{vector-eq}) for
plural F-strings for the right hand side under the inequality 
(\ref{vector-ineq}). Then possible configurations, which belong to 
the bound state of anti-quarks and quarks, are obtained for 
\beq
|\nu| N\geq 2\, ,
\eeq
which means the number of the anti-quarks is larger than two, so the total
number is larger than $N+3$. It is then larger than six for $N=3$. 
We give some example of these configurations below.

This condition is applied for the case of baryon, and we find for $N=3$
that the baryon configuration is obtained only for $\nu=0$ and $1$. In other words,
the three quarks ends only on the one cusp, namely there is no configurations
with separated distribution of quarks.

\vspace{.7cm}
\section{Numerical solutions for $|\nu| N= 2$}
\begin{figure}[htbp]
\vspace{.3cm}
\begin{center}
\includegraphics[width=14cm]{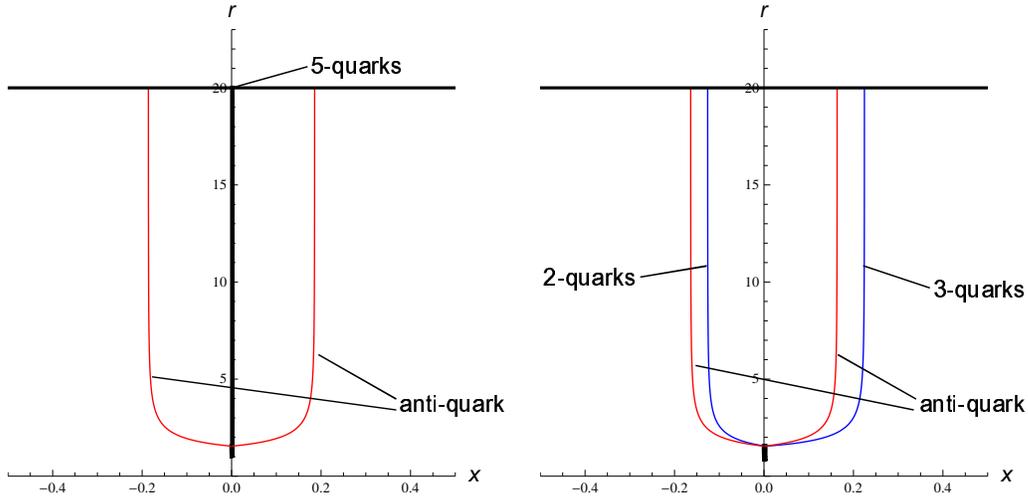}
\caption{\small Typical hepta-quark states for point D5 vertex. 
D5 brane, quarks and anti-quarks are respectively represented by 
the bold line at $x=0$, red curves and blue curves. Here, we take $N=3$,
 $q=0.3$, $\nu=-2/3$, $r(\theta=0)=1.54$ and the cutoff $r_{\rm{max}}=20$. In the left 
figure, two cusps stretch from $r(0)=1.54$ to $r_{\rm{max}}$ and five quarks 
flux emerge at the point $r_{\rm{max}}$ and the total energy $U_{\rm{tot}}$ is
 given by $U_{\rm{tot}}=142.0$.  In the right one, two cusps are 
almost localized at $r(\theta=0)=r(\theta=\pi)$ and $U_{\rm{tot}}=144.8$.  In the right, note 
that it seems that there are only two quarks but we just gather three 
quarks and two quarks together into one curve respectively.}
\label{point7}
\end{center}
\end{figure}
Here we restrict to the case of $N=3$ for the numerical calculations.
In this case, the penta quarks are unstable as seen above, then
we study here the configuration of the seven (hepta)-quark states, which
is the next simple multi-quarks state. We solve the equation of motion of
the system consisting of the D5 brane vertex and strings corresponding to
five quarks and two anti-quarks given by (\ref{U-total}) with the no force 
condition.  

\vspace{.7cm}
\noindent{\bf Point Vertex and Lowest Mass State:}

We consider firstly the simple vertex configuration, which is observed
as a point in our real three dimensional space. This configuration is 
obtained by solving (\ref{U-total}) by setting as $x=$ constant. Of course, we can 
choose at any point of $x$ where the vertex sits, we choose $x=0$ for  
simplicity.  Then the equation of motion for $r(\theta)$ is given as 
\cite{GI}
\beq
\partial_{\theta}\left({r^{\prime}\over\sqrt{r^2+(r^{\prime})^2}}\,
\sqrt{V_{\nu}(\theta)}\right)
-\left(1+{r\over 2}\partial_r\Phi\right){1\over\sqrt{r^2+(r^{\prime})^2}}\,
\sqrt{V_{\nu}(\theta)}=0~.
\eeq
The vertex extends from $r(0)(>0)$ to $r(\pi)$ with $\theta$, 
and the strings of the quarks and anti-quarks are attached at $r(\pi)$ and $r(0)$ 
respectively for negative $\nu(=-2/3)$.

\begin{figure}[htbp]
\vspace{.3cm}
\begin{center}
\includegraphics[width=7cm]{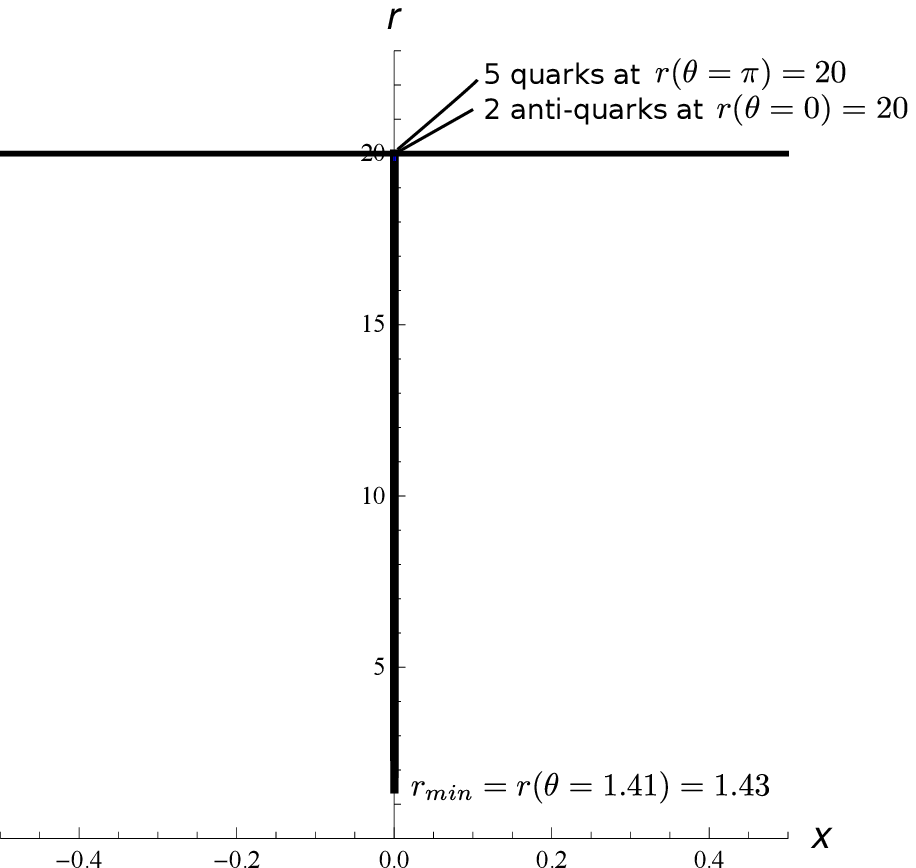}
\includegraphics[width=7cm,height=6.5cm]{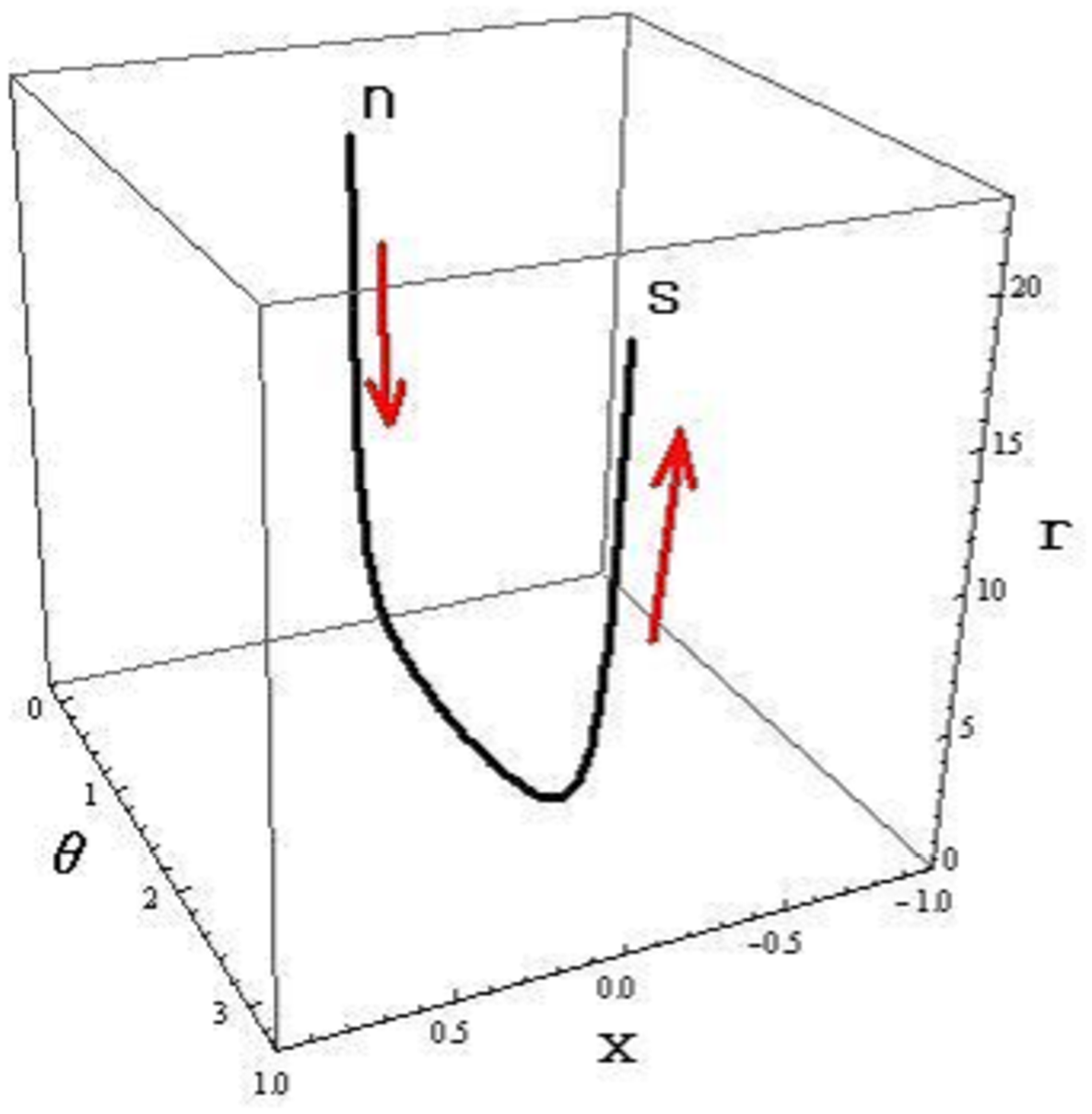}
\caption{\small The lowest energy configuration of point vertex.  
Here we take $N=3$, $q=0.3$, $\nu=-2/3$, $r(0)=r(\pi)=20$, so 
each cusp is at the boundary. We find this configuration has 
the lowest energy $U_{\rm{tot}}=140.7$. The configuration is given
in $x-r$ plane (left) and in $\theta -r- x$ three dimensional
 space (right). In the right figure, the arrows indicate the orientation
of the $U(1)$ flux in the D5 brane. And, at the point n (s) two anti-quarks
(five quarks) couple to the brane.}
\label{point-20}
\end{center}
\end{figure}

As for the no-force conditions, (\ref{UFr}), (\ref{UFx}), they are 
represented by noticing $N|\nu|=2$ as follows,
\beq
{2r'\over \sqrt{r^2+{r'}^2}}
   =\sum_{i=1}^{2}{r_x^{(i)}\over \sqrt{(r_x^{(i)})^2+(r/R)^4}}\, , 
\label{UFrp}
\eeq
for $r$ direction and 
\beq
  0=\sum_{i=1}^{2}{{x_r^{(i)}}\over{\sqrt{1+(r/R)^4(x_r^{(i)})^2}}}\, , 
\label{UFxp}
\eeq
for $x$ direction. In this case, the condition in the $x$ direction
is given only by the F-strings.  

Here we introduce the position of the flavor brane at $r=r_{max}$.  This plays 
a role of a cut off of the coordinate $r$ to obtain finite energy of the 
states with strings. 
Namely the quark strings here are connecting the D5 brane and the D7 brane. 
In order to give the explicit 
hepta quark solutions mentioned above, the equations of motion are solved 
numerically since it would be impossible to solve them analytically.
After solving them, we show two explicit configurations in the 
Fig.~\ref{point7}. 
In the right hand configuration, two cusps of the vertex are near at 
$r=1.54$. Namely the vertex starts from
one cusp at $r(\theta=0)=1.54$ and goes down slightly in the $r$-direction
toward $r_{\rm min}$, 
then goes up to the other cusp at $r(\theta=\pi)=1.54$.  
This configuration has rather large energy $U$. 
It is possible to obtain the configuration
with smaller energy than that of this by pushing up the one cusp point
$r(\theta=\pi)>1.54$ keeping the other side cusp position, 
$r(\theta=0)=1.54$. Then we find the left configuration of the Fig.~\ref{point7}
as a limit of this deformation.
This configuration has smaller energy than the one of the right one, because
the energy of strings is larger than the one of
the D5 brane in which the same number of strings are absorbed. 
Namely, the five quark strings
of the right configuration are absorbed by the D5 brane, which
stretches from $r=1.54$ to $r=r_{\rm max}$.

\vspace{.3cm}
Due to the same reason, we could find the minimum energy configuration by
pushing the other cusp up to $r=r_{\rm max}$. Actually we can see this
fact by the numerical estimation of the energy by varying $r(0)$, the 
results are shown in the Fig.~\ref{hepta4}.
Then the lowest energy
configuration is shown 
in the Fig. \ref{point-20}, and its energy is given as $U_{\rm{tot}}=140.7$
for $N=3$, $q=0.3$, $\nu=-2/3$, $r(0)=r(\pi)=20=r_{\rm max}$.
Then the lowest energy of hepta-quark state 
is obtained by the D5 brane only. Namely, all quark-strings 
are absorbed in the D5 brane in the case of the lowest energy state.

\vspace{.3cm}
We should notice the following point for the above solution
with lowest energy, which should be expected to be stable.
From the left figure of Fig.~\ref{point-20},
one may suspect an instability due to the
pair annihilation of quarks and anti-quarks
on the top of the vertex. However, such a possibility is avoided since
they are separated into the opposite poles, $n$ and $s$, 
of $S^5$ as shown in the right-figure, which is
given in the $x-r- \theta$ three dimensional space.
The distance between $n$ and $s$ is about $2 r_m$ in the $r$ direction.

Another apprehension is the instability due to the tachyon 
which appears when a D brane and an anti-D brane are facing with a 
small distance \cite{Sen}. The left figure looks like a similar configuration since
the oppositely oriented D brane parts are facing at the same point of $x$.
But, as seen from the right figure, it shows a five dimensional sphere 
with the radius which varies with $\theta$, and it is not bending as
being misunderstood from the left figure. Thus there is no reason to consider
the tachyonic instability.

\begin{figure}[htbp]
\vspace{.3cm}
\begin{center}
\includegraphics[width=6cm]{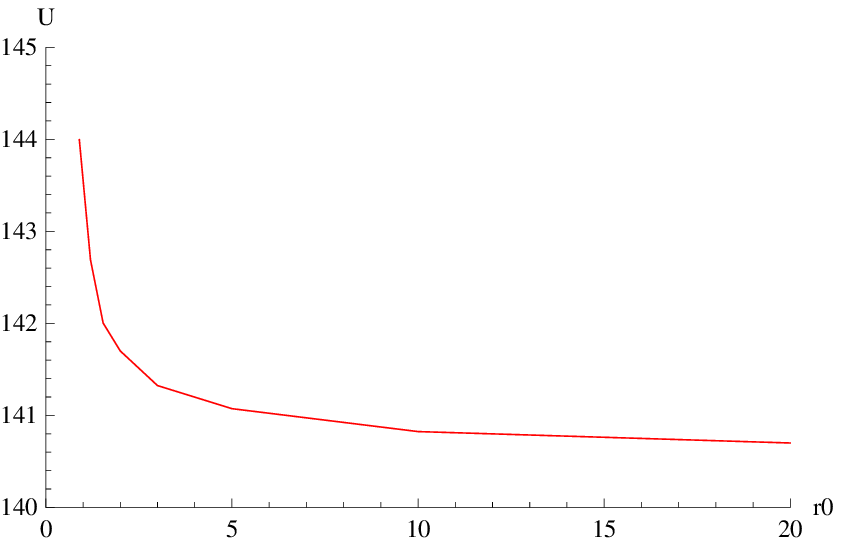}
\includegraphics[width=6cm]{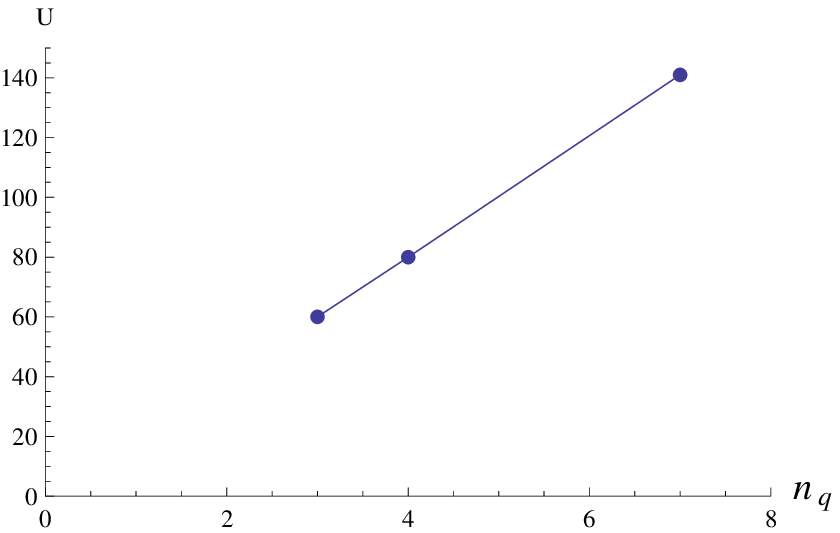}
\caption{\small Left;  The minimum of total energy of hepta-quark 
energy in the case of $L=0$ at $R=1$, $q=0.3$ and $r(\theta=\pi)=20$.  
The range of $r(0)$ is from 0.9 to 20.
Right; The lowest energy of multi-quark baryonic state for the various quark
number $n_q$ are given, $U=60(n_q=3), 80(n_q=4)$ and $141(n_q=7)$. }  
\label{hepta4}
\end{center}
\end{figure}
\vspace{.3cm}

\vspace{.3cm}

The situation is similar to the case of the baryon ($\nu=0,~1$). The minimum
energy of the baryon is also obtained by the configuration of D5 brane
only, which extends from $r_{\rm max}$ to $r_{\rm min}$ and absorbs three
quark strings for $N=3$ case. And its mass is about $60/R$ for $r_{\rm max}=20$
\cite{GI} for $N=3$. On the other hand, in the present case, the seven 
quark mass is about $141/R$ (see Fig.~\ref{hepta4}), 
which is about 2.4 times of the lowest baryon mass. Then the ratio 
$m_{\rm hepta}/m_{B}$ is well approximated by the quark number ratio, 
$7/3\sim 2.3$.  
This implies that the vertex energy is 
proportional to the quark number $n_q$, and it might be written as
\beq
 m=m_q^{\rm eff}n_q\, ,
\eeq
where $m_q^{\rm eff}$ represents the effective quark mass. Here
we can approximate as $m_q^{\rm eff}\propto r_{\rm max}-r_{\rm min}$,
We notice the above formula is satisfied when $r_{\rm min}$ is preserved
as a constant for any state of $n_q$.
In this case, it suggests a quark counting rule of masses of 
baryonic states. In the right of Fig. \ref{hepta4}, we show this relation
by adding the baryonium state with the quark number $n_q=4$,
and its mass is obtained as 80/R \cite{GINT2,GINT3}.
We expect that this rule would hold for other hadronic states. 
In the below, we show that the rule 
also holds for the ones of split vertex.

\vspace{.7cm}
\noindent{\bf Split Vertex and its Tension}
\begin{figure}[htbp]
\vspace{.3cm}
\begin{center}
\includegraphics[width=14cm]{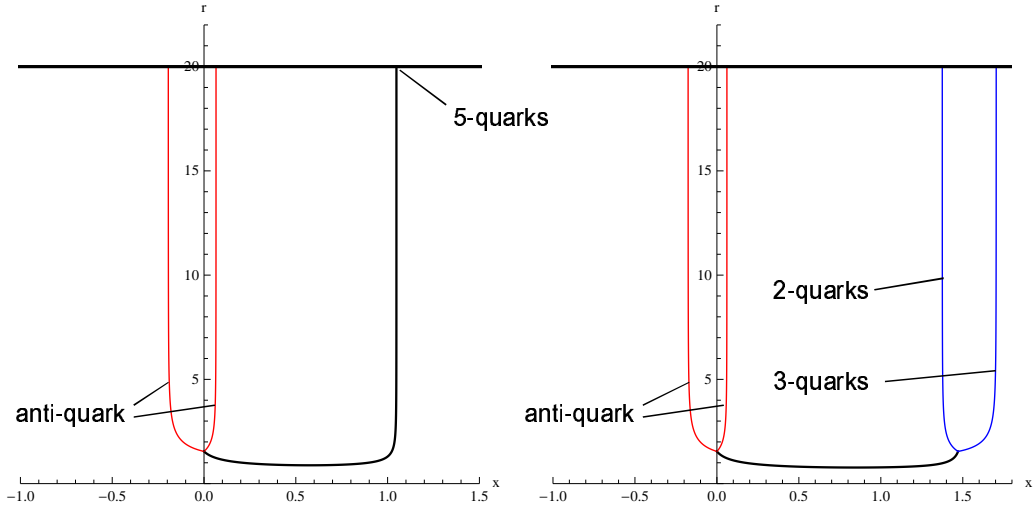}
\caption{\small Typical split vertex hepta-quarks for various $r(\theta=\pi)$ 
at $N=3$, $q=0.3$, $R=1$ and $r(\theta=0)=1.54$. In this case, D5 brane (bold 
curve) also stretches to $x$-direction.  Two curves 
correspond to the point vertex case in Fig.~\ref{point7}. In the left
 one, the configuration has $L=1.05$ and $U_{\rm{tot}}=142.6$. In the right
 one, the configuration has $L=1.47$ and $U_{\rm{tot}}=145.5$. 
\label{hepta-quarks}}
\end{center}
\end{figure}
\vspace{.3cm}
\vspace{.7cm}
\begin{figure}[htbp]
\vspace{.3cm}
\begin{center}
\includegraphics[width=6cm]{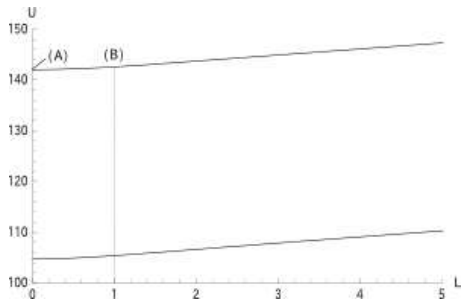}
\caption{\small The upper curve shows total energy of split hepta-quarks 
and the lower curve shows the vertex energy of split hepta-quarks in the 
case of $R=1$, $q=0.3$, $r(0)=1.54$ and $r(\pi)=20$.  
$L$ represents the length of the split of D5 brane, namely 
$L=|x(\theta=0)-x(\theta=\pi)|$. The point A and B correspond to the energy
of the left of Fig.\ref{point7} ($L=0$) and Fig. \ref{hepta-quarks} 
($L=1$)respectively.}  
\label{hepta-energy}
\end{center}
\end{figure}
\vspace{.3cm}

Next, we consider higher energy state by extending the vertex in the 
our three dimensional space, say $x$ direction. Such a
configuration is called as the ``split vertex'' of
D5 brane similar to the baryon case \cite{GI}. Here we concentrate
on the tension of this extended D5 vertex.

For this configuration, the no force conditions 
are given by (\ref{UFr}), 
(\ref{UFx}) and we solve the equations of motion (\ref{U-total}) or
(\ref{Heom})$\sim$(\ref{Heom3}) in Appendix without fixing
$x(\theta)$. Therefore, unlike the case of point vertex, split vertex solutions have a configuration of 
length of vertex separation $L$ defined by 
$L=|x(\theta=0)-x(\theta=\pi)|$.  
  

In the Fig.~\ref{hepta-quarks},  it is shown two typical examples of the
split hepta-quark state which satisfy no force conditions and
the equation of motion, and each figure corresponds to 
the ones of point vertex in Fig.~\ref{point7}.
The left one of Fig.~\ref{hepta-quarks} represents split vertex 
configuration of 
hepta-quarks which starts from the cusp $(x(0),r(0))=(0,1.54)$ 
where two anti-quarks couple, to the cusp 
$(x(\pi),r(\pi))=(1.05,20)$ where five quarks couple. 
We can say the similar statement about the right one of 
Fig.~\ref{hepta-quarks}. 


\vspace{.3cm}
We find that
the value of the total energy is higher than that of point vertex by the
configuration of vertex length $L$.
Then the energy difference would be about 
\beq
 \Delta U\simeq \tau_{\rm D5}L
\eeq
where
$\tau_{\rm D5}$ denotes the tension of the D5 brane and is given below.
In the Fig.~\ref{hepta-energy}, we show the relation of the total energy and
vertex energy as a function of $L$ for $r(\pi)=20$ and a fixed $r(0)=1.54$.
The total energy increases with the vertex 
separation $L$, and its minimum is given at 
$L=0$, namely for the point vertex as expected.

The two curves are almost parallel, then the slope of them represents
the tension of the vertex only. It seems 
to converge to $2\sqrt{q}/R^2$, which is twice of one fundamental string.
Actually,
in numerical analyses of Fig.~\ref{hepta-energy}, at $L=5$ the one for 
upper curve is $2.18\sqrt{q}/R^2$, while its value of lower one is 
$2.21\sqrt{q}/R^2$. Moreover, at $L=400$ the both slope is equal to
  $2.01\sqrt{q}/R^2$. Therefore, we can confirm that in the 
limit $L\to\infty$, the slope of the energy seems to converge 
to $2\sqrt{q}/R^2$ as mentioned above.

We know that the tension of D5 brane at each cusp is 
estimated by the one of the string times their number at the
cusp point. 
In this case, the number of anti-quarks of cusp at $\theta=0$ is two,
then the tension obtained above represents the brane tension at $\theta=0$.
The reason that the point $\theta=0$ of the brane is pulled is that
the tension at $\theta=\pi$ is larger than the one at $\theta=0$ due to
the number of the strings attached at the cusp. This implies that
the point with the smallest tension in the brane is stretched when we
pull the brane.


\vspace{.3cm}

\section{Penta-quark in other confining theories }\label{penta}

In the previous sections, we show that the five-quark state, which is made
of one anti-quark and $N+1$ quarks, cannot be constructed in our model given
by Eqs.(\ref{finite-c-sol}) and (\ref{dilaton}), and we need two anti-quarks
are needed at least to form a penta-quark like exotic state. Here we show that
this conclusion is common to other holographic confining gauge theories
through two typical examples.

\subsection{Non-Susy solution}
For, the non-supersymmetric case, the solution is obtained as \cite{GY},
\beq
ds_{10}^2=e^{\Phi_{\rm NS}/2}\left(\frac{r^2}{R^2}A^2(r)\eta_{\mu\nu}dx^{\mu}dx^{\nu}+\frac{R^2}{r^2}dr^2+R^2d\Omega_5^2
\right),
\eeq
where
\beq
A(r)=\left(1-\left(\frac{r_0}{r}\right)^8\right)^{1/4},\quad
e^{\Phi_{\rm NS}}=\left(\frac{(r/r_0)^4+1}{(r/r_0)^4-1}\right)^{\sqrt{3/2}}.
\eeq
This configuration has a singularity at $r=r_0$. 
So we cannot extend our analysis to near the singularity where 
higher curvature contributions are important. This theory 
provides confinement and chiral symmetry breaking at zero 
temperature. The confinement is realized due to the gauge 
condensate, which is proportional to $r_0^4$ in the present 
case. The chiral symmetry breaking means that the massless 
quark has the non-zero chiral condensate in this theory. 
Therefore, a dynamical quark mass would be generated for a 
massless quark. This point is different from the supersymmetric 
case.

As in the supersymmeric case, the D5 brane action (DBI term plus 
WZW term) is written as
\beq
S_{\rm NS}=T_5\Omega_4R^4\int dtd\theta\sin^4{\theta} 
           \left(-\sqrt{G(\theta)-\tilde{F}_{t\theta}^2}
           +4\tilde{A}_t(\theta)\right)\, ,
\eeq
\beq 
G(\theta)=e^{\Phi_{\rm NS}}A(r)^2(r^2+r^{\prime 2}
          +(r/R)^4 A(r)^2 x^{\prime 2})\, .  
\eeq
From the equation of motion for $\tilde{A}_t$, we obtain 
\beq
\partial_{\theta}\left(\frac{\sin{\theta}^4\tilde{F}_{t\theta}}
{\sqrt{G(\theta)-\tilde{F}_{t\theta}^2}}\right)
=-4\sin^4{\theta}=\partial_{\theta}D\, ,
\label{d'}
\eeq
where $D$ is the same one given in (\ref{D3}) and (\ref{d}).
The action is thus rewritten by eliminating the gauge field 
in terms of the Legendre transformation as follows,
\beq
U_{\rm NS}
=\frac{N}{3\pi^2\alpha^{\prime}}\int d\theta\sqrt{G(\theta)V_{\nu}(\theta)}\, ,
\qquad
V_{\nu}(\theta)=D(\nu,\theta)^2+\sin^8{\theta}\, .  
\eeq

\noindent{\bf No-force condition}

As in the supersymmetric case, we can find the no-force condition 
between the system of D5 brane and (anti-)quarks. 
The total energy of the system is represented as follows,
\beq
U_{\rm total}=U_{\rm NS}+U_{\rm F-q}+U_{\rm F-anti-q},
\eeq
where
\bea
&&U_{\rm F-q}
=\sum^{N(1+|\nu|)}_{i=1}\frac{1}{2\pi\alpha^{\prime}}
\int_{r^{(i)}(\theta=\pi)}^{r_{\rm max}}dr^{(i)}A(r^{(i)})
e^{\Phi_{\rm NS}^{(i)}/2}\sqrt{1+\left(\frac{r^{(i)}}{R}\right)^4A
(r^{(i)})^2x_r^{(i)2}}\, ,\\
&&U_{\rm F-anti-q}
=\sum^{N|\nu|}_{i=1}\frac{1}{2\pi\alpha^{\prime}}
\int^{r_{\rm max}}_{r^{(i)}(\theta=0)}dr^{(i)}A(r^{(i)})
e^{\Phi_{\rm NS}^{(i)}/2}\sqrt{1+\left(\frac{r^{(i)}}{R}\right)^4
A(r^{(i)})^2x_r^{(i)2}}.
\eea
At the cusp $\theta=0(\theta=\pi)$, we give the balance of the tensions
between the D5 brane and the $|\nu|N((1+|\nu|)N)$ F-strings corresponding 
to the anti-quarks (quarks).  Then, the no-force conditions are given for 
$\theta=0$ as for $r$ direction,
\beq
N|\nu|\frac{r'}{\sqrt{r^2+r'^2+(r/R)^4A(r)^2x'^2}}
=\sum^{N|\nu|}_{i=1}\frac{r_x^{(i)}}{\sqrt{r_x^{(i) 2}+(r/R)^4A(r)^2}}\, . 
\eeq
For $x$ direction as,
\beq
N|\nu|\frac{x'}{\sqrt{r^2+r'^2+(r/R)^4A(r)^2x'^2}}
=\sum^{N|\nu|}_{i=1}\frac{x_r^{(i)}}{\sqrt{1+(r/R)^4A(r)^2 x_r^{(i)2}}}\, .
\eeq
At the boundary $\theta=\pi$, we obtain the condition with the same 
form by replacing $N|\nu|$ by $N(1+|\nu|)$ in the above conditions.  

As a special case of baryon, let consider the configuration of 
$|\nu|N=1$,  $N=3$ and $\nu<0$, namely penta-quark. By considering 
the above condition at $\theta=0$, we obtain
\beq
r^2=0.
\eeq
In the present case, we must consider the region of $r$ as
$r\geq r_0>0$.
Thus, the penta-quark cannot be realized by this baryon model in 
the non-supersymmetric and confining theory because of the 
existence of the singularity at $r=r_0$.  But the hepta-quark 
state ($N|\nu|=2$) is also allowed for this model.  And other 
states with $N|\nu|>2$ is also possible.  

\subsection{Type IIA D4/D4 Model} 

In type IIA model, the baryon vertex is given by D4 brane, which wraps on
$S^4$ in the background of stacked D4 \cite{cgst,SSu}.  
In this model, the 10d background metric is given by 
\bea
&&ds_{10}^2=\left(\frac{r}{R}\right)^{3/2}\left[f(r)d\tau^2+
\eta_{\mu\nu}dx^\mu dx^\nu\right]
+\left(\frac{R}{r}\right)^{3/2}f(r)^{-1}dr^2+R^{3/2}{r}^{1/2}
d\Omega_4^2\, , \\
&&f(r)=1-r_{h}^3/r^3, \qquad R^{3}=\pi\gs N \ls^{3}, \qquad
r_{h}=\frac{16\pi^2}{9} R^{3} T^{2}\, , 
\label{d4metric}
\eea
And, the world-volume action of D4 brane is given as
\beq
\label{action1}
S = -T_4 \int d^5\xi e^{-\tilde{\phi}}\sqrt{-\det(g+\tilde F)}
  +T_4 \int A_{(1)}\wedge G_{(4)}~, 
\eeq
\beq
e^{\tilde\phi}
=\left({r\over R}\right)^{3/4},
\eeq
where $T_4=1/(g_{s}(2\pi)^{4}\ls^{5})$, and the world-volume gauge field 
$A_{(1)}$ couples to the dual of the background four-form field strength 
$G_{(4)}$\, .

Using the explicit background\req{d4metric} one can rewrite the action 
in the form 
\beq
S = - T_4\Omega_3 R^3\int dt d\theta\,\left[\sin^3\theta\sqrt{G_{(4)}-\tilde F_{t\theta}^2}+\tilde F_{t\theta}D_{(4)}\right]\, ,
\label{action2}
\eeq
\beq
 G_{(4)}=r^2+f(r)^{-1}r^{\prime 2} +(r/R)^3(x^{\prime 2}+f(r)\tau'^2)\, ,
\label{D4}
\eeq
where $\Omega_{3}=2\pi^2$.  
In the above expression (\ref{D4}), we added the freedom $\tau(\theta)$. 
However, we do not solve
the classical equations of motion for it in the present 
paper.  The displacement $D_{(4)}$ now satisfy the equation 
$$
\partial_{\theta}D_{(4)} =-3\sin^3\theta\, ,
$$
and is consequently given by 
\beq
D_{(4)}=3\cos\theta-\cos^3\theta-2+4\nu~\, .
\label{d4}
\eeq
The constant of integration has been written again in terms of a
parameter $\nu<0$. 

$\tilde F_{t\theta}$ and $D_{(4)}$ is related by the equation of 
motion for $\tilde A_{t}$ as, 
\beq
\tilde F_{t\theta}={{D_{(4)}\sin^3\theta}\over{\sqrt{D^2_{(4)}+\sin^6\theta}}}
\sqrt{G_{(4)}}\, .
\label{D}
\eeq
Then the action is rewritten as, 
\bea
S &=&  -T_4 \Omega_{3}R^3\int dt d\theta
\sqrt{D_{(4)}}\sqrt{V_\nu}\, ,\\
V_\nu &=& D^2_{(4)}+\sin^6\theta\, ,
\label{action3}
\eea

\noindent{\bf No-force condition}

No-force condition in this model is obtained as similar way in D5-model.

The system is represented by the following total energy, 
\beq
  U_{\rm total}={N\over 8\pi\alpha'}\int_0^{\pi} d\theta~\sqrt{G_{(4)}} 
  \sqrt{V_{\nu}(\theta)}~+U_{\rm F-q}+U_{\rm F-anti-q}\, ,
\eeq
where 
\beq
 U_{\rm F-anti-q}=\sum_{i=1}^{N|\nu|}{1\over{2\pi\alpha'}}
  \int^{r_{\rm max}}_{r^{(i)}(\theta=0)}dr^{(i)}~\sqrt{D_{s}^{(i)}}\, ,
\eeq
\beq
  U_{\rm F-q}=\sum_{i=1}^{N(1+|\nu|)}{1\over{2\pi\alpha'}}
  \int^{r_{\rm max}}_{r^{(i)}(\theta=\pi)}dr^{(i)}~\sqrt{D_{s}^{(i)}}\, ,  
\eeq
\beq
 D_{s}^{(i)}=f(r^{(i)})^{-1}
+(r^{(i)}/R)^{3}((x_r^{(i)})^2+f(r^{(i)})(\tau_r^{(i)})^2)\, ,
\eeq
and $x_r=\partial_r x$.

At the cusp $\theta=0$ ($\theta=\pi$), we give the balance of the 
tensions between the brane and the $|\nu| N$ ($(1+|\nu|) N$) F-strings 
corresponding to the anti-quarks (quarks).  They are given for $\theta=0$ 
as, 
\beq
N|\nu|{r'\over \sqrt{G_{(4)}}} 
   =\sum_{i=1}^{N|\nu|}{r_x^{(i)}\over \sqrt{D_{s}^{(i)}}} \, ,
\label{UFr4}
\eeq
for $r$ direction
\beq
N|\nu|{x'\over \sqrt{G_{(4)}}} 
   =\sum_{i=1}^{N|\nu|}{{x_r^{(i)}}\over{\sqrt{D_{s}^{(i)}}}}\, ,
\label{UFx4}
\eeq
for $x$ direction
\beq
N|\nu|{\tau'\over \sqrt{G_{(4)}}}
 =\sum_{i=1}^{N|\nu|}{\tau_x^{(i)}\over \sqrt{D_{s}^{(i)}}}\, ,
\label{UFtau4}
\eeq
and for $\tau$ direction.  At the boundary $\theta=\pi$, we obtain 
the conditions with the same form by replacing
$|\nu| N$ by $(1+|\nu|) N$ in the above equations.
 
These no-force conditions imply $r^2=0$ as in D5-model when $N|\nu|=1$.
However, $r$ is restricted to the region $r\geq r_h>0$.  
Thus the penta quark state is forbidden as in D5-model when $N|\nu|=1$.  
But the hepta-quark state ($N|\nu|=2$) is allowed.  And other states 
with $N|\nu|>2$ is also possible.

\vspace{2cm}
\section{Summary and Discussion}\label{comparison}

New type of holographic baryonic-states are studied by solving the system of 
the D5 brane and F-strings. They are embedded in a 10D background, 
which is dual to a confining gauge theory. 
The D5 brane wraps on $S^5$ and couples to the self-dual five 
form field strength formed by the $N$ stacked D3 branes. And they 
are dissolved into the D5 brane as the $U(1)$ electric field. 
This electric field could come out from two poles on the $S^5$ 
and they are replaced outside of D5 brane by 
the F-strings corresponding to quarks. For the usual baryon, $N$ 
quarks ($N$ number of fluxes) go out from the cusps.

The embedded solution 
is characterized by an integral constant $\nu$, which determines the 
distribution of the outgoing fluxes from the D5 brane and also the 
orientation of the flux.  
Choosing negative $\nu$, 
we find negative number ($-N|\nu|$) of flux at 
one pole, and positive number of flux on the other pole is given by $N(1+|\nu|)$, 
which is larger than the total number of the quarks given by $N$. However, these 
configuration has the same total flux number with the 
one of the baryon since
they are constructed with anti-quarks and quarks.

The quarks and the anti-quarks are bounded at the opposite side of the 
two cusps of the vertex separately, so they cannot vanish through the pair 
annihilation of quark and anti-quark on the vertex since the two cusps 
on D5 brane are separated in the 10d space-time.  

Simple example of such a state is considered in the case of $N=3$ (like QCD) 
and $\nu=-1/3$.  This corresponds to the so-called penta-quark state 
($4Q+1\bar Q$), but this state is not realized because of the no-force 
conditions for $\bar Q$. 
In the case of baryon, this implies that the allowed configurations are
obtained only for $\nu=0$ and $1$.

 On the other hand, the state of $\nu=-2/3$ 
has $2\bar Q $ on one cusp of the vertex, then the no-force condition 
could be satisfied and this hepta-quark state ($5Q+2\bar Q$) is allowed.  
The states with larger number of anti-quarks on the one cusp of the 
vertex are possible, but they have larger mass.  Here we studied the 
configurations and the masses of the hepta-quark states by fixing as 
$N=3$ for simplicity. We find the lowest mass is proportional to the number
of quarks, namely, the mass of the hepta quark is about 7/3 times of the baryon
mass. 

The excitation of the state is realized by extending the 
quark strings or the D5 vertex. In the case of the former, the energy 
would be needed proportional to the number of the extending strings 
times a string tension, which is here given by $\sqrt{q}/R^2$.  For 
the extension of the D5 brane, we observed that its tension is about 
$2\sqrt{q}/R^2$ for hepta quark. This implies that the extending part of the vertex 
is near the cusp point where two anti-quarks couple to.  

We also could assure 
that the qualitative situation obtained for the supersymmetric background 
is also seen in the other 
confining theories considered here, namely for Non-Susy D5 theory and Type IIA 
D4/D4 model.

\vspace{.5cm}
Finally we briefly comment on the other possible model  
to form the penta-quark by introducing 
two D5 and one anti-D5 branes.  The two D5 branes are combined to 
the anti-D5 brane by two anti-quark strings, and four quarks and 
one anti-quark are connected to the two D5 branes and one anti-D5 
brane, respectively.  It 
would be an interesting problem to estimate its mass and compare it 
with the one of the hepta quark state obtained here.  This would be 
done in the next step of our work.  


\vspace{.5cm}
\section*{Acknowledgements}
The authors like to thank M. Ishihara for useful discussions, and 
T. Taminato wishes to thank all the members of High Energy group
of Kyushu University.

%
\appendix
\section*{Appendix; Equations of motion with parameter $s$}

\vspace{.23cm}

Here we show another formulation of solving the equations of motion 
derived from (\ref{u}) used in \cite{cgst,GI}.  
Firstly, rewrite (\ref{u}) in terms of a 
general world-volume parameter $s$ defined by
functions $\theta=\theta(s)$, $r=r(s)$, $x=x(s)$ as: 
\beq
U = {N\over 3\pi^2\alpha'}\int ds~e^{\Phi/2}
\sqrt{ r^2\dot{\theta}^2 + \dot{r}^2+(r/R)^{4}\dot{x}^2}~
\sqrt{V_{\nu}(\theta)}\, ,
\label{upar2}
\eeq
where dots denote derivatives with respect to $s$.  
Then the momenta conjugate to $r$, $\theta$ and $x$ are given as 
\beq
p_r=\dot{r}\Delta, \quad
p_{\theta}=r^2\dot{\theta}\Delta, \quad
p_{x}=(r/R)^{4}\dot{x}\Delta, \quad
\Delta=e^{\Phi/2}\frac{\sqrt{V_{\nu}(\theta)}}
   {\sqrt{ r^2\dot{\theta}^2 + \dot{r}^2+(r/R)^{4}\dot{x}^2}}~\, .
\label{mom}
\eeq
Since the Hamiltonian that follows from the action\req{upar2} vanishes  
identically due to reparametrization invariance in $s$.  Then we 
consider the following identity 
\beq
2\tilde{H} =p_r^2 + \frac{p_{\theta}^2 }{r^2}+\frac{R^4}{r^4}p_{x}^2-
\left(V_{\nu}(\theta) \right) e^{\Phi}=0~\, .
\label{ham}
\eeq
Regarding this constraint as a new Hamiltonian, we obtain 
the following canonical equations of motion, 
\bea\label{Heom}
\dot r &=& p_r~,~~ \dot p_r =\frac{2}{r^5}p_x^2 R^4
     + \frac{p_\theta^2}{r^3}+{1\over 2}
        \left(V_{\nu}(\theta)\right)~ e^{\Phi}\partial_r\Phi,\\
\dot\theta &=&\frac{p_\theta}{r^2}~, ~~\dot p_\theta = -6\sin^4\theta \left(\pi\nu-\theta+ 
\sin\theta\cos\theta
\right)~e^{\Phi}, \\
\dot x &=& \frac{R^{4}}{r^{4}}p_{x},~~\dot{p}_{x}= 0
\label{Heom3}
\eea
The initial conditions should be chosen such that $\tilde H=0$.  
By solving these equations, we could find the same solutions given above.


\end{document}